# Time-resolved fast neutron imaging: simulation of detector performance

**D. Vartsky,[a] I. Mor,[a] M. B. Goldberg,[a] I. Mardor,[a] G. Feldman,[a] D. Bar,[a] A. Shor,[a] V. Dangendorf,[b] G. Laczko,[b] A. Breskin,[c] R. Chechik[c]**

*[a]Soreq NRC, Yavne, Israel*

*[b]Physikalisch-Technische Bundesanstalt, Braunschweig, Germany*

*[c]Weizmann Institute of Science, Rehovot, Israel*

*** Corresponding author:**

D. Vartsky, Tel: 972-8-9434589; Fax: 972-8-9434676; Email: david@soreq.gov.il

**Abstract**

We have analyzed and compared the performance of two novel fast-neutron imaging methods with time-of-flight spectroscopy capability (see V. Dangendorf et al., these Proceedings). Using MCNP and GEANT code simulations of neutron and charged-particle transport in the detectors, key parameters such as detection efficiency, the amount of energy deposited in the converter and the spatial resolution of both detector variants have been evaluated.

## 1. Introduction

Over the last two years we have developed and tested two variants of time-resolved, fast-neutron imaging detectors, that combine sub-mm resolution spatial imaging with energy spectroscopy [1,2]. A detailed description of the two systems has been presented by Dangendorf et al [2,3]. Briefly, the first of these, a Fast Neutron Gas-filled (FANGAS) detector [3], consists of a hydrogenous neutron/proton converter (polypropylene) coupled to a triple-step Gas Electron Multiplier (GEM). Position information is obtained by the signal induced through a resistive anode on a pad-structured pickup electrode.

The second, an Optical Fast-Neutron Imaging detector (OTIFANTI) [2], consists of a fast plastic scintillating screen, viewed by a CCD camera through suitable optics and a gated image-intensifier.

In the present work we report on a study, based on MCNP [4] and GEANT [5] computer-simulation codes, that evaluated several key intrinsic properties of the neutron converters in the two variants, namely: detection efficiency, amount of energy converted into a measurable signal and position resolution.

## 2. FANGAS detector

The neutron converter in this detector is a sheet of polypropylene of density 0.92 g/cm$^3$. Fast neutrons interact with the converter primarily by producing knock-on protons. Some of the protons escape from the converter and are detected by the adjacent charged-particle gas detector. The converter sheet thickness is determined by a compromise between efficient neutron absorption and high escape probability of protons from the converter into the gas.

### *2.1. Detection efficiency and proton energy spectrum*

The calculation of detection efficiency vs. sheet thickness was performed using the GEANT code. In it, we calculated the number of protons escaping the converter into the gas per incident fast-neutron and assumed that each escaping proton is detected. In addition, the proton energy spectrum was determined.

Fig. 1 shows the calculated detection efficiency vs. converter thickness for three incident neutron energies: 2 MeV, 7.5 MeV and 14 MeV.

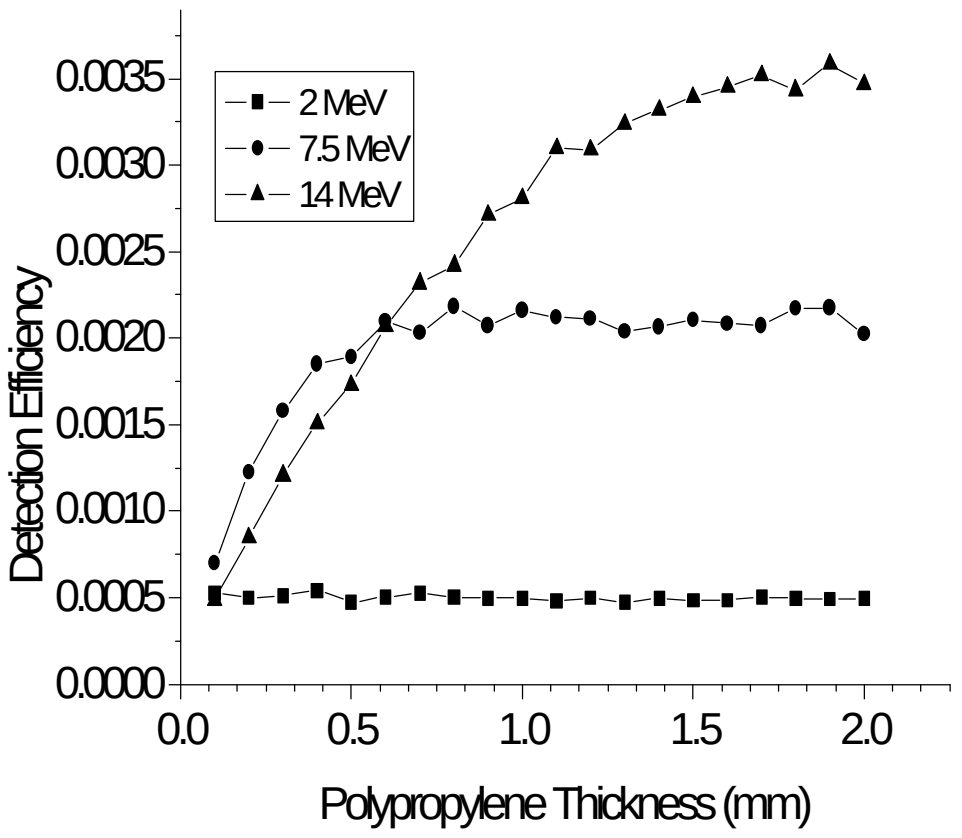

Fig. 1. FANGAS detection efficiency vs. converter thickness.

As evident from Fig. 1, detection efficiency initially increases with converter thickness, but then tends to saturate when the thickness approaches the proton range in the material. The efficiency is rather low: ~0.2% for a 1-mm-thick converter at a neutron energy of 7.5 MeV. To overcome this problem it is envisaged to stack ~25 detector elements, thus attaining reasonable detection efficiencies of ~ 5%.

Fig. 2 shows the energy spectra of protons emitted from the converter at the same three neutron energies, calculated for an angular range of 0-27$^o$ relative to the incident neutron direction. This angular range corresponds to an angle subtended by a 0.5x0.5 mm$^2$ detection pixel located at a distance of 0.5 mm below the converter [3].

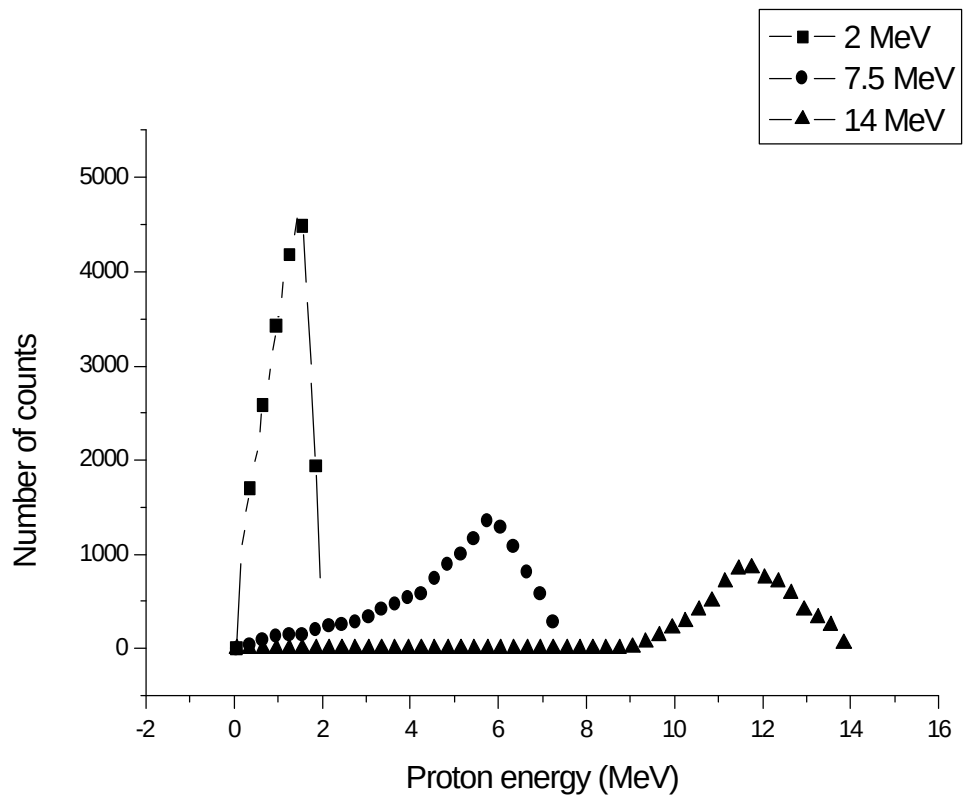

Fig. 2. Energy spectra of protons reaching a 0.5x0.5 mm$^2$ detection pixel situated at a distance of 0.5 mm from the converter

The emitted proton energy distributions are rather broad and centred on average energies of 1.14, 4.97 and 11.7 MeV for 2, 7.5 and 14 MeV neutrons, respectively. Spectra of protons emitted at larger angles (pixels at a larger lateral distance from the point of interaction) will be softer.

### *2.2. Influence of neutron scattering on contrast*

As mentioned above, attaining detection efficiencies of ~5 percent will require stacking ~25 individual detectors. Thus, it was necessary to study the extent of image contrast deterioration caused by neutrons scattered into a detector by its neighbours.

For this purpose, we have performed MCNP calculations of an image of 4 carbon blocks located at a distance of 300 cm from a stack of 25 polypropylene sheets, 1 mm thick, spaced at 9 mm from each other. Polypropylene sheet dimensions were 200x200x1 $mm^3$ and they were divided into 1x1 $mm^2$ pixels. The neutron source was a point source positioned at 6 meters from the detector. The incident neutron energy was 7.75 MeV. On the basis of published cross-sections, the expected contrast of the carbon blocks at this energy is 20%. For each detector sheet, the number of neutrons arriving at each pixel and their energy spectrum were calculated.

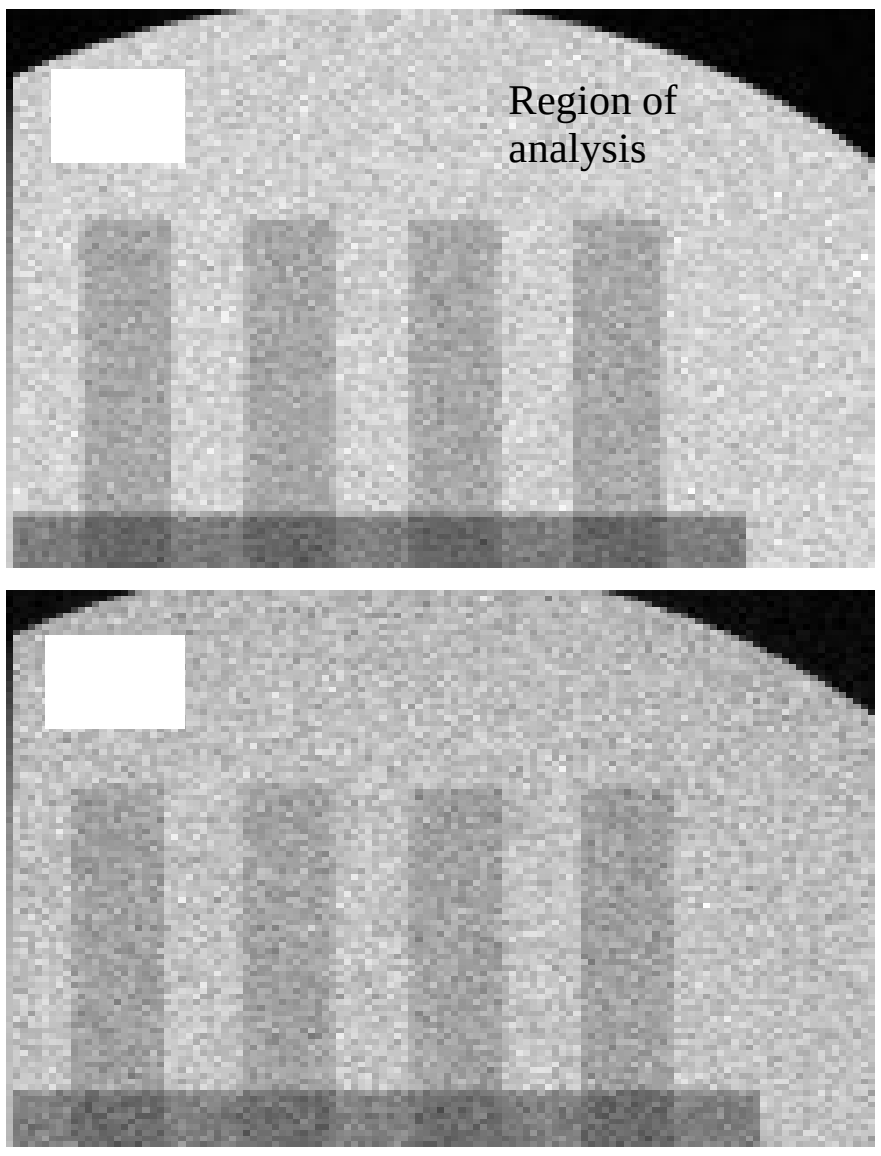

Fig. 3. Simulated neutron images of 4 carbon blocks at the 1$^{st}$ (top) and 25$^{th}$ (bottom) converter sheets of a FANGAS stack detector

Fig. 3 shows the computer-simulated neutron images obtained with the 1$^{st}$ and 25$^{th}$ polypropylene converter sheets. The average number of counts/pixel was 310±18 and 217±16 for these sheets, respectively.

Fig. 4 shows profiles taken across the simulated C-blocks images for the 1$^{st}$ and 25$^{th}$ sheet. As can be seen, the contrast at the 25$^{th}$ sheet decreases from the expected 20% to 16.5%, due to neutron scattering.

Fig. 5 shows the energy spectra of neutrons arriving at the1$^{st}$ and 25$^{th}$ sheets. Three features are visible in the spectra, on top of a flat continuum due to scattering from hydrogen. The peak on the right is the portion of the 7.75 MeV incident neutron beam which carries the radiographic information. The middle and left peaks are broad structures due, respectively, to elastic and inelastic scattering of neutrons by C present in neighbouring sheets. As can be seen, the contributions of the latter reactions increase with sheet number.

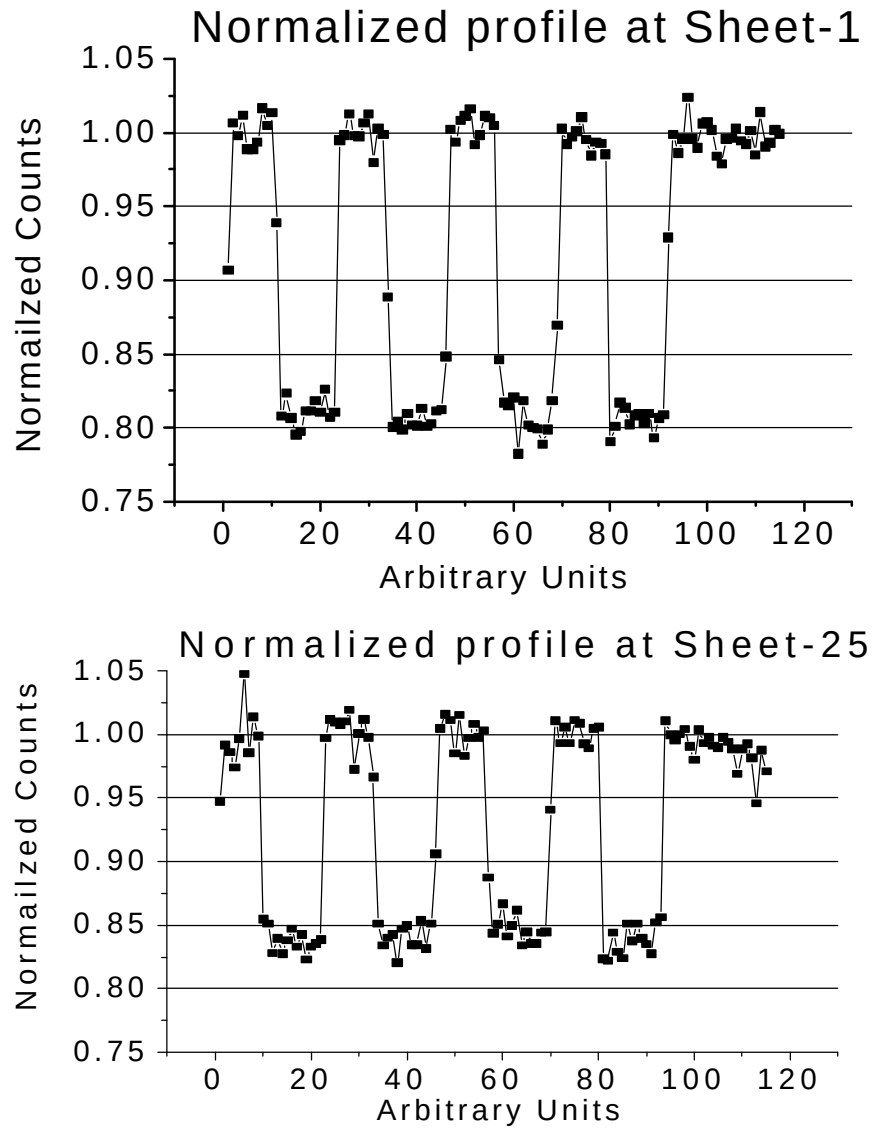

Fig. 4. Profile of the C-blocks image for the 1$^{st}$ and 25$^{th}$ sheet

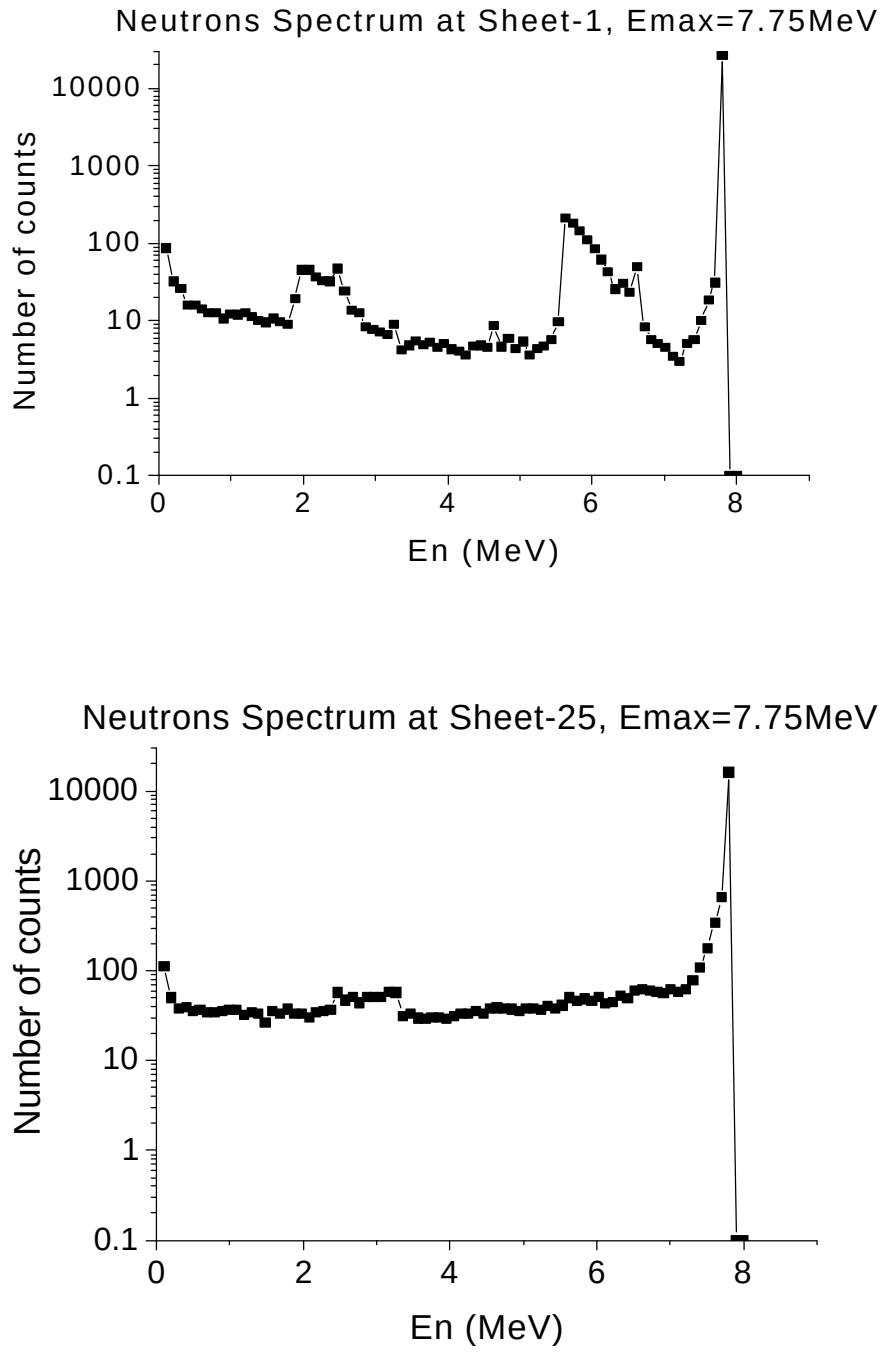

Fig. 5. Neutron energy spectra at the 1st and 25th sheet

Fig 6 shows the simulated variation of the number of scattered neutrons with. sheet number, as well as the ratio of transmitted to scattered neutrons. As can be observed, the number of scattered neutrons initially increases until it reaches a maximum around sheet number 13. From this point on, it decreases slowly, due the self absorption of the incident flux.

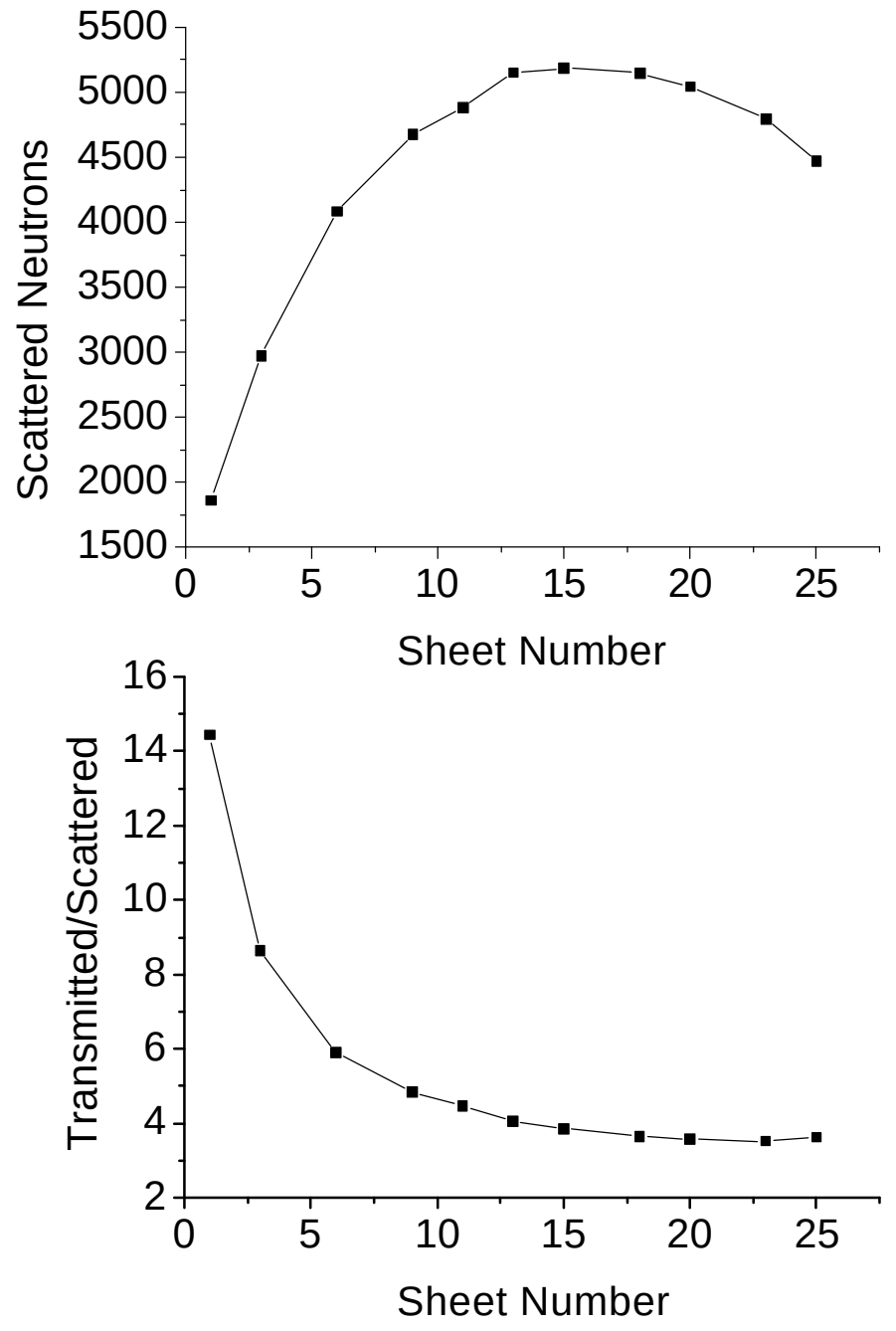

Fig. 6. The number of scattered neutrons (top) and the ratio of transmitted to scattered neutrons (bottom) as function of converter sheet number.

The transmitted-to-scattered neutron intensity ratio decreases by a factor of 4 from sheet number 1 to 25.

### *2.3. Position resolution*

The position resolution of the FANGAS detector has been determined in a GEANT simulation. First, we calculated the point spread function (PSF) for a single sheet of 1 mm thick polypropylene converter. An infinitesimally-narrow incident neutron beam normal to the polypropylene converter irradiates it at a single point. The position distribution of protons emitted from the converter is determined at a distance of 0.5 mm below the converter, which corresponds to the halfway distance between it and the first GEM [3].

Fig. 7 shows the PSF obtained in this calculation for incident neutron energies of 2, 7.5 and 14 MeV. As can be observed, the PSF becomes progressively broader with increasing neutron energy. This reflects the phenomenon that, as the initial neutron energy increases, so does the angular spread of knock-on protons emerging from a converter sheet into the gas.

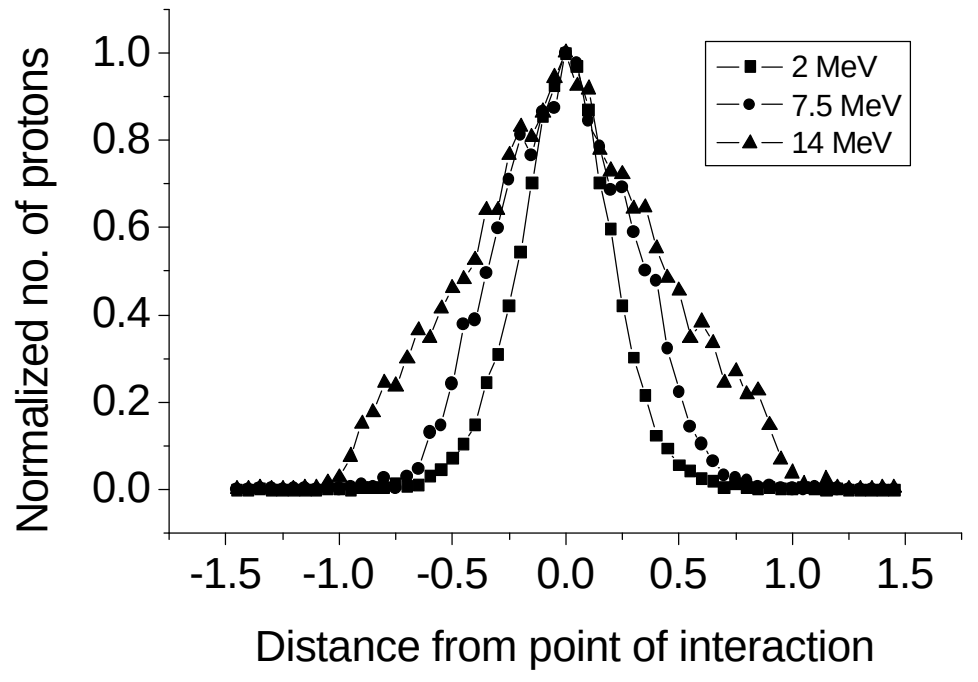

Fig. 7. Point spread function of a single FANGAS detector module

Table 1 lists the calculated FWHM and FWTM of the FANGAS PSF.

Table 1. FWHM and FWTM of the FANGAS point spread function

| Neutron Energy | FWHM (μm) | FWTM (μm) |
|---|---|---|
| 2.0 MeV | 440 | 900 |
| 7.5 MeV | 700 | 1200 |
| 14.0 MeV | 900 | 1800 |

## 3. OTIFANTI detector

In the present work, the OTIFANTI detector consists of a fast-plastic (polystyrene) scintillator fiber screen, 200x200x20 $mm^3$ in dimensions. The reason for using scintillating optical fibers, rather than a plain plastic scintillator slab, is to maintain the position resolution, independent of screen thickness. The minimal fiber dimensions in these calculations were 50x50 $\mu m^2$. A neutron interacts in the fibers and transfers part of its energy to a proton, which produces scintillations in the fiber. A fraction of this light travels along the fiber and is emitted at its ends. The light is transported via a front-coated mirror and large-aperture lens to an optical detection system. In the present work, the performance of the screen was characterized by calculating neutron conversion efficiencies, using tabulated cross-sections for H and C. The GEANT code was used to determine emitted light yields and the position resolution.

### *3.1 Neutron conversion efficiency and light emission*

The conversion efficiency, i.e., the probability of creating a knock-on proton (from hydrogen) per incident neutron, was calculated using known cross-sections for reactions on hydrogen and carbon. The ratio of H/C in polystyrene is 1 and the density is 1.05 g/cc[6]. Fig. 8 shows the detection efficiency of the polystyrene scintillator vs. neutron energy for detector thicknesses of 10, 20, 50 and 100 mm. The fluctuations visible, especially for the thicker screens, are due to resonances in the carbon cross-section.

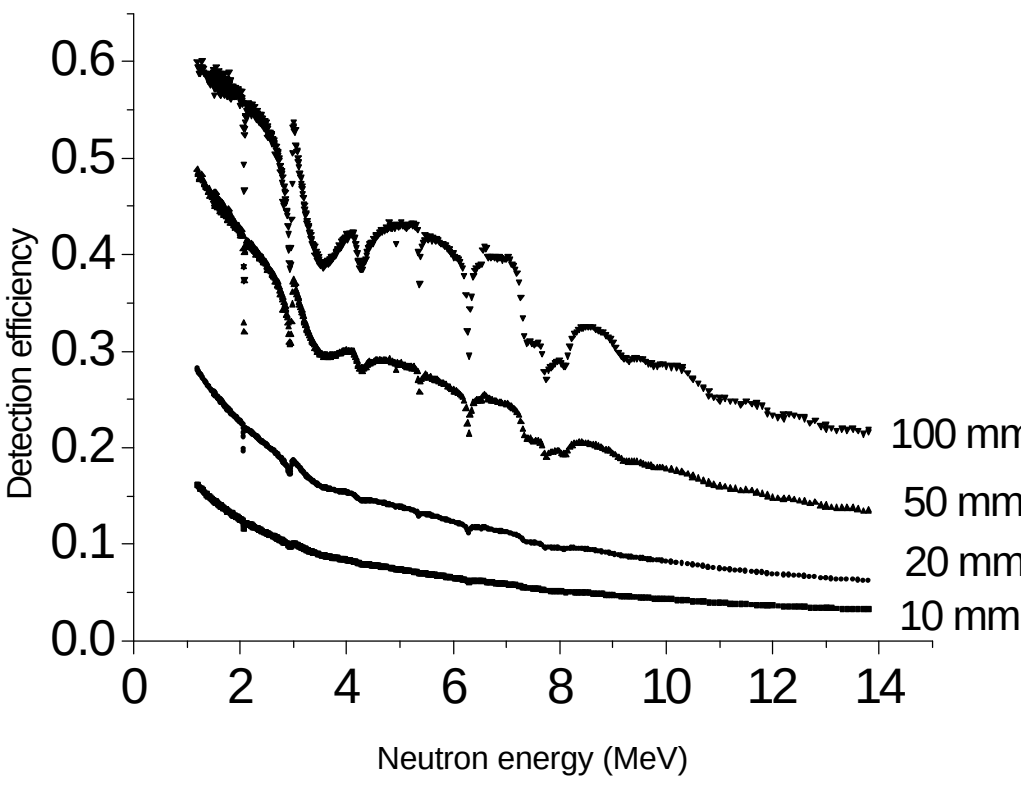

Fig. 8. Detection efficiency of OTIFANTI detector vs. neutron energy for various detector thicknesses

The calculated detection efficiency of a 20 mm thick screen is about 10% at 7.5 MeV.

In order to determine the distribution of energy deposited in the fibers and the amount of light created by the neutrons, we have performed a GEANT simulation of the energy deposited by the protons in a 0.5x0.5 $mm^2$ pixel of a fiber screen, 200x200x20 $mm^3$ in dimensions. In the simulation, this screen was irradiated uniformly by neutrons incident normal to its face. Fig. 9 shows the distribution of the energy deposited by the protons in the fibers. The average proton energy is 0.78, 2.44 and 3.22 MeV for the 2, 7.5 and 14 MeV neutrons, respectively.

Using the known entrance and exit energies of each proton in the fiber we have also calculated the number of light photons created in the fiber by the incident neutrons, using the non-linear proton-energy to-light conversion relation for a plastic scintillator [7]. The average amounts of light created in the fiber were 2600, 10,600 and 17,400 light photons for 2, 7.5 and 14 MeV neutrons, respectively.

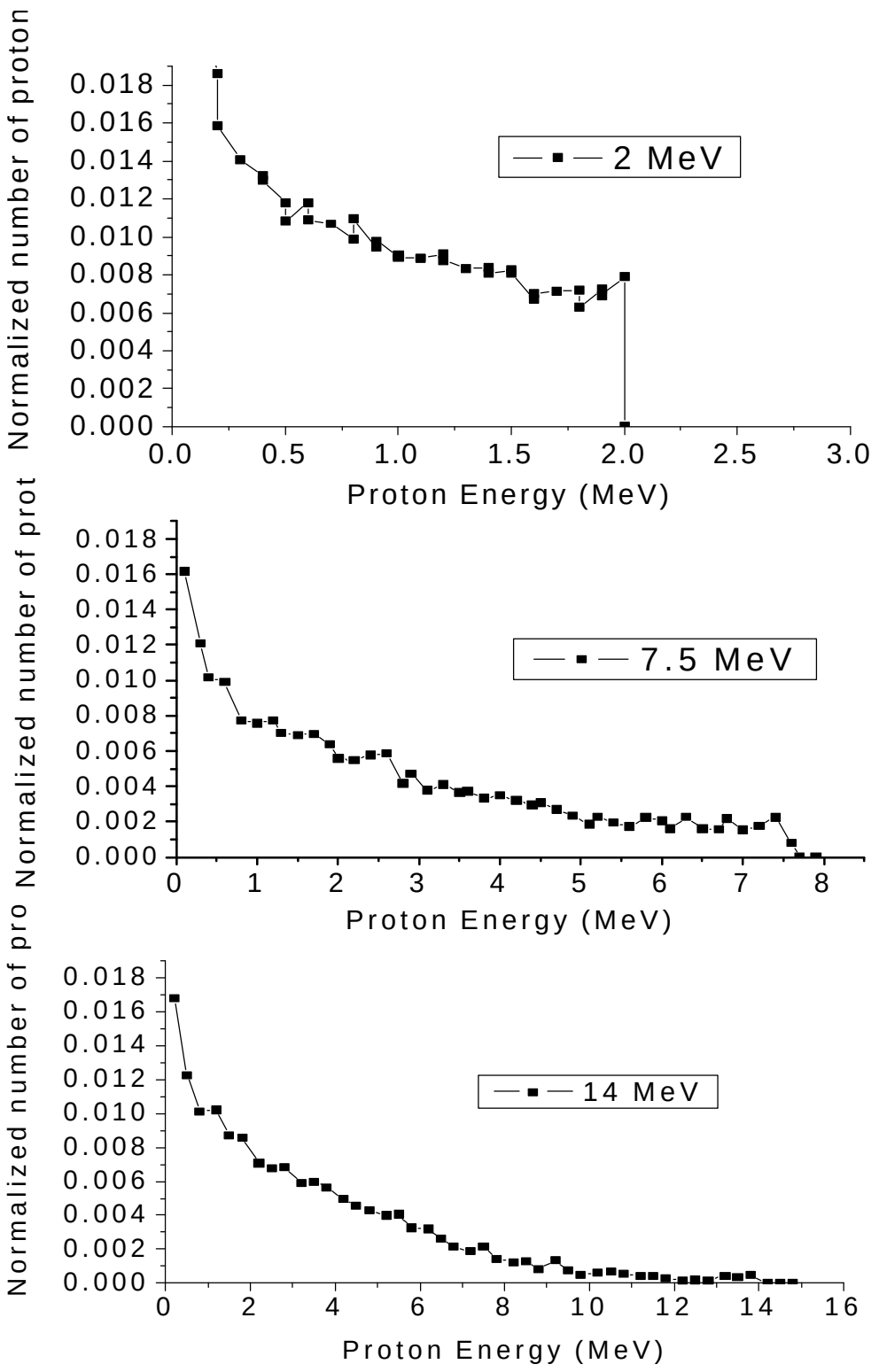

Fig. 9. Spectra of energy deposited by protons in 0.5x0.5 $mm^2$ fibers for 2 (top), 7.5 (middle) and 14 MeV (bottom) neutrons. The number of protons was normalized to the number of incident neutrons/pixel

### *3.2 Position resolution*

The position resolution of the scintillating fiber screen was determined using GEANT. In these calculations we determined the point spread function by exposing the screen to an infinitesimally thin neutron beam, incident normal to its surface. The fiber dimensions were 50x50 ìm$^2$. The amount of light created in the irradiated fiber and those adjacent to it fibers was determined. The light in the neighbouring fibers is created by protons crossing from the irradiated fiber to its neighbours and by scattered neutrons. Possible cross-talk of light from fiber to fiber was not taken into account in these calculations.

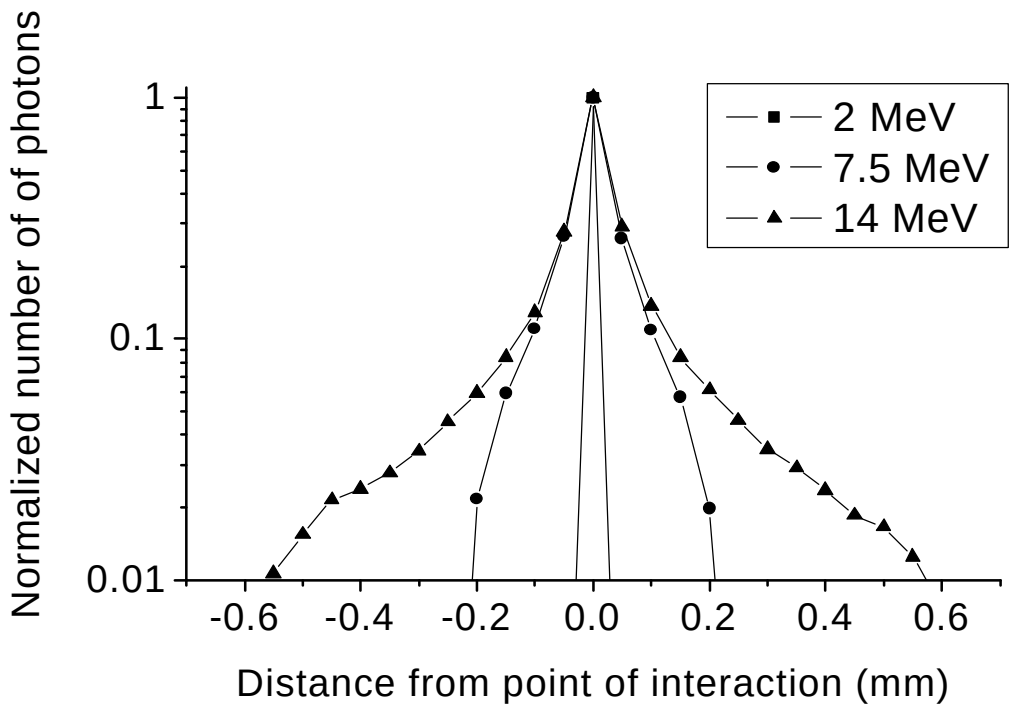

2

Fig. 10 shows the point spread function of the fiber screen. As expected, it becomes broader as the neutron energy increases.

Table 2 lists the simulation-derived FWHM and FWTM values of the above distributions.

Table 2. FWHM and FWTM of OTIFANTI point spread function

| Neutron Energy | FWHM (ìm) | FWTM (ìm) |
|---|---|---|
| 2.0 MeV | ~60 | 80 |
| 7.5 MeV | 90 | 200 |
| 14.0 MeV | 100 | 300 |

Lastly, the modulation transfer function (MTF) of the screen was determined by simulating an image of a carbon penetrameter, the number of line-pairs/mm ranging from 0.5 to 5.

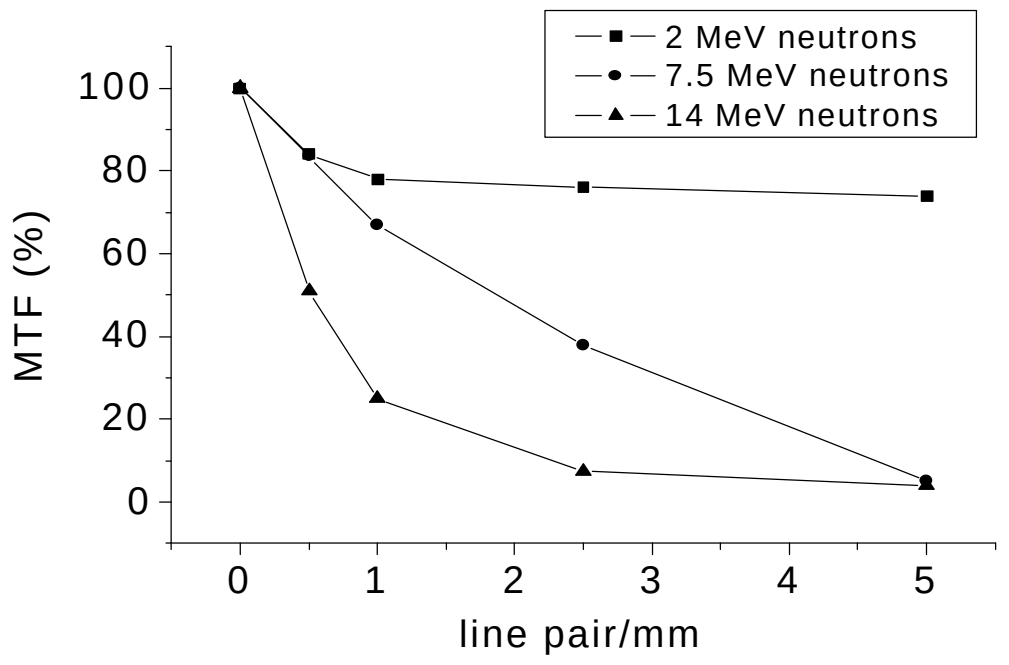

Fig. 11. MTF of the fiber screen. Fiber dimensions 50x50ìm$^2$

Fig. 11 shows the calculated MTF of the fiber screen. The MTF of the 2 MeV neutrons is excellent, due to the very short range of the protons. As expected, it deteriorates as the neutron energy increases.

## 4. Conclusions

The performance of neutron converters for two fast neutron imaging detector variants was analyzed using Monte Carlo methods. In order to achieve a neutron efficiency comparable with the OTIFANTI variant, a significant number of detector units of the FANGAS variant would have to be stacked.

Moreover, the intrinsic position resolution of OTIFANTI appears to be significantly better than that of FANGAS. However, the cross-talk of light between the fibers in OTIFANTI is likely to degrade the position resolution somewhat.

V.D. acknowledges the support of the Minerva Foundation. A.B. is the W.P. Reuther Professor of Research in the Peaceful Use of Atomic Energy